\documentclass{ieeeaccess}
\usepackage[T1]{fontenc}
\usepackage{cite}
\usepackage{amsmath,amssymb,amsfonts}
\usepackage{algpseudocode}
\usepackage{algorithm}
\usepackage{graphicx}
\usepackage{multirow}
\usepackage{booktabs}
\usepackage{textcomp}
\usepackage{hyperref}
\usepackage{mathrsfs}
\usepackage{float} 
\usepackage{bm}
\def\BibTeX{{\rm B\kern-.05em{\sc i\kern-.025em b}\kern-.08em
    T\kern-.1667em\lower.7ex\hbox{E}\kern-.125emX}}
\begin{document}
\history{}
\doi{}

\title{Quantum Patches: Enhancing Robustness of Quantum Machine Learning Models}
\author{\uppercase{Ban Q. Tran}\authorrefmark{1,2,3}, \IEEEmembership{Student, IEEE}, \uppercase{Chuong K. Luong}\authorrefmark{3} , \uppercase{Viet Q. Nguyen}\authorrefmark{3}, \uppercase{Duong M. Chu}\authorrefmark{3}, and {Susan Mengel}\authorrefmark{1},
\IEEEmembership{Member, IEEE}}
\address[1]{Department of Computer Science, Texas Tech University, TX 79416 USA (email: bantran@ttu.edu)}
\address[2]{Department of Computing Fundamental, FPT University, Hanoi, 10000, Vietnam (email: bantq3@fe.edu.vn)}
\address[3]{SAP Innovation Lab, FPT University, Hanoi, 10000, Vietnam}


\markboth
{Tran \headeretal: Quantum Patches: Enhancing Robustness of Quantum Machine Learning Models}
{Tran \headeretal: Quantum Patches: Enhancing Robustness of Quantum Machine Learning Models}

\corresp{Corresponding author: Ban Q. Tran (email: bantran@ttu.edu).}

\begin{abstract}
Machine learning models and their applications, such as autonomous driving systems, are becoming increasingly common and are essential components of human daily life. However, due to their sensitivity to perturbed noise, these models are easily susceptible to adversarial attacks. Not only are classical machine learning models affected, but quantum machine learning (QML) models have also been proven to be vulnerable to adversarial attacks, which degrade their performance. To defend against these types of attacks, several classical methods have been proposed. Among these, a prominent approach uses various types of pseudo-noise during training to enhance the model's robustness against real-world attacks. One of the recently emerging solutions is to leverage the unique properties of quantum circuits to create quantum-based pseudo-noise similar to real perturbed noise to counter adversarial attacks. This paper proposes a solution that utilizes random quantum circuits (RQCs) as adversarial data to help QML models overcome these adversarial attacks. The results reported in this paper show that the data generated by RQC actually provides a similar effect to models trained with adversarial data on high-feature datasets. This quantum-based pseudo-noise resulted in a significant reduction in the attack rate in the CIFAR-10 data set, from \textbf{89. 8\%} to \textbf{68.45\%}. For the CINIC-10 dataset, the successful attack rate decreased from \textbf{94.23\%} to \textbf{78.68\%}. This research opens up avenues for applying unique quantum properties, such as superposition, entanglement, and even decoherence, to enhance the quality of machine learning models.
\end{abstract}

\begin{keywords}
 Quantum Machine Learning, Quantum Deep Learning, Quantum Neural Networks, Quantum Adversarial Machine Learning, Adversarial Attacks, Variational Quantum Algorithms, Parameterized Quantum Circuits, Random Quantum Circuits.
\end{keywords}

\titlepgskip=-15pt

\maketitle

\section{Introduction}
\label{sec:introduction}
\PARstart{A}{}dversarial attacks \cite{b1}  have become widespread in recent years and seriously threaten machine learning and deep learning models \cite{b2}. Some previous studies have shown that machine learning models are highly vulnerable to carefully crafted perturbations of input data, leading to misclassifications even though these perturbed images are barely distinguishable to the human eye \cite{b3}. In theory, QML models have been shown to outperform classical machine learning models in processing speed \cite{b4}. However, research also indicates that QML is similarly vulnerable to adversarial attacks, just like its counterpart \cite{b5}. Specifically, if you take a clean image of a panda, add a small amount of carefully crafted noise, and pass it through a quantum classifier, the model will misclassify it as an image of a gibbon. The primary cause of this phenomenon is the linear nature of QML models \cite{b6}. These adversarial attacks could have catastrophic consequences for humans when deep learning models are widely deployed in real-world applications, such as self-driving taxis or medical robots. Therefore, comprehensive research into solutions against these attacks is necessary to protect machine learning and deep learning models.
\Figure[t!](topskip=0pt, botskip=0pt, midskip=0pt)[width=1.4\columnwidth]{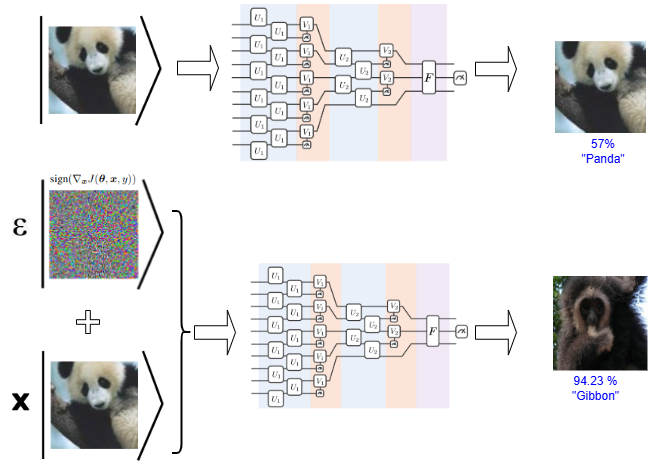}
{Simulation of the adversarial attack mechanism, in which the original image, a panda, is perturbed with epsilon noise to deceive the model. The human eye can easily filter out the noise and classify the image correctly. However, the model is misled as a gibbon due to the multidimensional noise amplification in the input data.\label{AA}}

Quantum computing \cite{b31} has conferred quantum benefits over classical algorithms by its distinctive principles and phenomena like superposition and entanglement. The quantum advantages mainly arise from implementing single and multi-qubit quantum gates and manipulating quantum states. RQCs are quantum circuits capable of randomly generating various single and multi-qubit gates with an unspecified circuit depth \cite{b32}. The functioning of RQCs results in data transformation driven by the quantum kernel within the Hilbert space. This transformation has been widely utilized in data augmentation techniques, allowing QML models to enhance generalization and improve learning quality. This paper focuses on leveraging data augmentation techniques \cite{b11} to help QML models defend against adversarial attacks. By exploiting the unique properties of RQCs in transforming spatially-local data space, we demonstrate that RQCs significantly improve the robustness of QML models against this attack.
\section{RELATED WORKS}
Various categories of adversarial attacks include white-box attacks, black-box attacks, targeted attacks, untargeted attacks, one-step attacks, and iterative attacks \cite{b15}. These attack methods fundamentally depend on techniques that provide adversarial instances to mislead clasical neural networks (CLNN) or classical deep learning (CDL) models. Notable techniques include the Fast Gradient Sign Method (FGSM) \cite{b3}, Projected Gradient Descent (PGD) \cite{b16}, Basic Iterative Method (BIM) \cite{b11}, DeepFool \cite{b35}, and Momentum Iterative Method (MIM) \cite{b18}. There are several defense methods to help deep learning models against adversarial attacks, such as adversarial training \cite{b3, b19}, adversarial example detection \cite{b8}, input preprocessing \cite{b9}, gradient masking \cite{b10, b24}, data augmentation \cite{b11}, robust model design \cite{b12}, ensemble of defenses \cite{b13}, defensive distillation \cite{b25}, defense-GAN \cite{b26}, and explainability research \cite{b14}. These techniques have all, to varying degrees, been shown to be effective in improving model robustness. Fundamentally, most techniques aimed at enhancing the robustness of CLNN models can also be applied to QNN models, as the nature of QNN architectures is similar to that of CLNN, differing only in the design of the neural network components. However, in this review section, we primarily focus on studies that leverage quantum properties to improve the robustness of QNN models.

Several studies have discussed methods to help QNN models defend against adversarial attacks. In summary, three main key defensive techniques are currently being used to improve adversarial robustness for QNN models: adversarial robustness through quantum noise, certifiable robustness of quantum classifiers, and adversarial training. Quantum noise is an undesirable phenomenon that occurs when qubits undergo decoherence during the processing of a quantum computer. When building quantum computers, it is essential to minimize the occurrence of quantum noise, allowing the quantum computer to perform more efficiently. However, the inherent nature of quantum noise has the effect of mitigating adversarial attacks. The concept of utilizing quantum noise has been explored in several studies and shown potential results \cite{cohen2019certified, du2021quantum, weber2021optimal}. Specifically, depolarization noise in a quantum circuit reduces quantum differential privacy \cite{zhou2017differential}. It indirectly limits the possible influence of adversarial perturbations, while still preserving the usefulness of the quantum classifier due to its inherent robustness to depolarization noise. The second key technique utilizes certifiable robustness \cite{weber2021optimal}, which employs quantum hypothesis testing (QHT) \cite{helstrom1967detection, holevo1973statistical} to achieve provable robustness. The core idea of QHT is to ensure that an optimal measurement cannot reliably distinguish two quantum states. If this condition is met, the quantum classifier will treat the output quantum states of a clean image and an adversarial example as identical, assigning them the same label. The study demonstrated that the robustness limit they found is theoretically optimal in the case of binary classification and when the probability of the classifier's top-choice label is greater than $1/2$ \cite{weber2021optimal}. The final and most common key technique is adversarial training, in which adversarial examples are generated during the model training process and included in the training sets \cite{b3, b11, wong2020fast, bai2021recent}. Several studies have demonstrated the effectiveness of this method for quantum classifiers \cite{b5, ren2022experimental}. For instance, one study conducted adversarial training on the MNIST dataset and achieved 92\% accuracy after training for approximately 40 epochs on both legitimate and adversarial test samples \cite{b5}. This result stands in complete contrast to vanilla training, where the model was easily fooled and the accuracy dropped to zero after only 4–5 iterations. However, one issue with this method is that the model's training and validation processes typically use the same type of adversarial attack noise. This problem raises the question of whether models, having been trained using adversarial examples that do not match the new attack type, can still maintain their robustness and defend against these different types of adversarial examples if the attacker changes their approach \cite{kang2019transfer}. Arising from this question, we proposed a method utilizing RQC and leveraging quantum properties to generate a type of "pseudo-adversarial examples." The goal is to create a more generalized adversarial example and apply the adversarial training method to enhance the resilience of QNN models.


\section{THEORETICAL BACKGROUND}

\subsection{Adversarial Attacks}
\Figure[t!](topskip=0pt, botskip=0pt, midskip=0pt)[width=0.9\columnwidth]{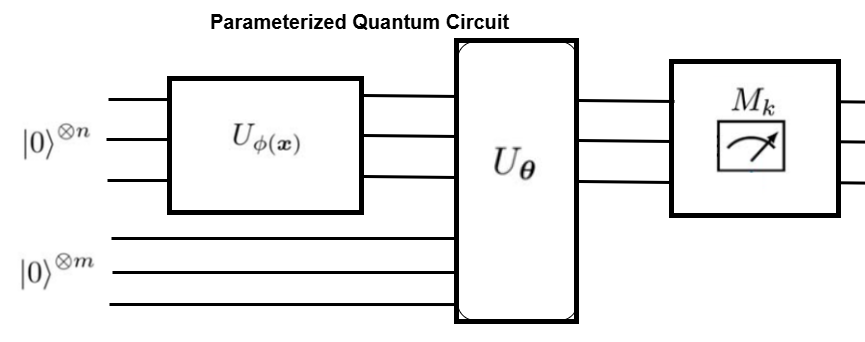}
{The abstract architecture of a typical PQC in Hilbert space.\label{PQC}}
Adversarial attacks are small, carefully crafted perturbations inserted by hackers into input data to confuse machine learning models \cite{b1}. Although these subtle disruptions are often imperceptible to the human eye, they pose significant challenges for machine learning models \cite{b1}. Figure \ref{AA} illustrates the mechanism of an adversarial attack, where noise is added to the input image, causing the model to misclassify it as a "Panda" instead of a "Gibbon". Similar to classical machine learning, the mechanism of adversarial attacks in QML also employs the same principle. Specifically, a small perturbation $\delta \in \nabla$, is introduced into the input sample $x=\left( x_{1},\ldots ,x_{A}\right)$ to optimize the loss function $\mathcal{L}$.
\begin{equation}\delta \equiv argmax_{\delta^{'} \in \Delta}\mathcal{L}\left( f\left( x+\delta^{'};\theta ^{\ast }\right) ,y\right)\label{eq1}\end{equation}
where $f:\mathbb{R} ^{A}\rightarrow \mathbb{R} ^{B}$ is a QML model built based on parameterized quantum circuits (PQCs) that transforms input data $x$ in the vector space $A$ into softmax probability distributions of $B$ classes. A PQC is a lattice of quantum circuits where the horizontal dimension represents the qubits and the vertical dimension represents special quantum rotation gates that transform the quantum state. This transformation process can be adjusted via the parameters passed to these rotation gates. Fig. \ref{PQC} illustrates a typical PQC architecture. The noise added to the input data must not exceed a threshold value for a successful attack, for example $\Delta =\left\{ \delta \in \mathbb{R} ^{A}:\left\| \delta \right\| _{\infty }\leq \varepsilon \right\}$ where $\epsilon$ represents the perturbation strength of the attack and is typically below 0.1.
The essence of the issue with adversarial examples confusing both classical machine learning and QML models is that the precision of an individual input feature is limited \cite{b3}. This limitation makes it impossible for the classifier to distinguish between the input $x$ and the adversarial input $\begin{aligned} \overline{x}=x+\eta \end{aligned}$. Let's provide an example of the FGSM method for generating adversarial examples to attack a neural network. If $\theta$ represents the model's parameters, $x$ is the model's input, and $y$ is the target output corresponding to $x$, then $J(\theta, x, y)$ is the cost function used to train the model. The formula following can express the perturbation added to the input data.
\begin{equation}\eta =\varepsilon sign\left( \nabla _{x}J\left( \theta ,x,y\right) \right)\label{eq2}\end{equation}
The research has shown that with $\epsilon = 0.25$, a classical shallow softmax classifier can achieve an error rate of up to 99.9$\%$ on the MNIST dataset. $\epsilon = 0.1$ can result in an error rate of up to 87.15$\%$ for convolutional maxout network on the CIFAR-10 dataset \cite{b3}.
\subsection{QUANTUM COMPUTING}
In this section, we will present some of the fundamental building blocks that constitute the special properties of quantum circuits.
\Figure[t!](topskip=0pt, botskip=0pt, midskip=0pt)[width=0.9\columnwidth]{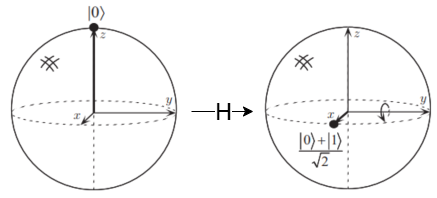}
{The Hadamard gate's mechanism for transforming the quantum state on a single qubit is described. This gate rotates the state vector from the z-axis to the y-axis. This mechanism creates the principle of superposition in quantum mechanics.\label{Had}}
\subsubsection{QUBITS}
\Figure[t!](topskip=0pt, botskip=0pt, midskip=0pt)[width=0.7\columnwidth]{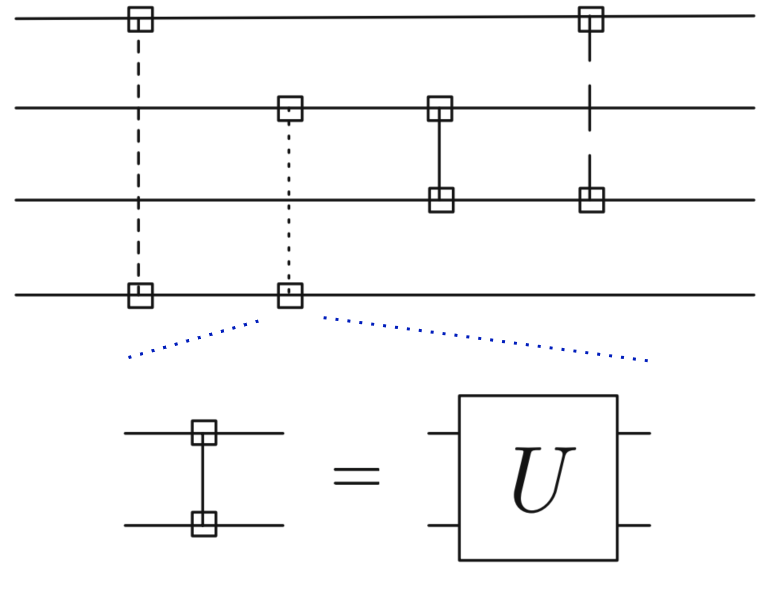}
{The architecture of a basic RQC takes as input a quantum state embedded from classical data. Unitary gates ($U$) are applied according to a random distribution to pairs of qubits to create extraordinary quantum phenomena, such as entanglement, and to transform the input quantum state into the corresponding output quantum states. The output quantum state will be collapsed into classical bitstrings using the measurement tool.  \cite{b39}. \label{RQC}}
A quantum bit, or qubit for short, is a unit of information in quantum computing, similar to the bits 0 or 1 in classical computing \cite{b36}. A qubit can exist in the state $| 0 \rangle$ or $| 1 \rangle$ and, uniquely, can also be in a linear combination of states, commonly referred to as superposition. The Dirac representation of a qubit takes the form $| \psi \rangle = \alpha | 0 \rangle + \beta | 1 \rangle$, where $| 0 \rangle$ and $| 1 \rangle$ are computational basis states, and $\alpha$ and $\beta$ are amplitudes, typically in the form of complex numbers.
\Figure[t!](topskip=0pt, botskip=0pt, midskip=0pt)[width=2.0\columnwidth]{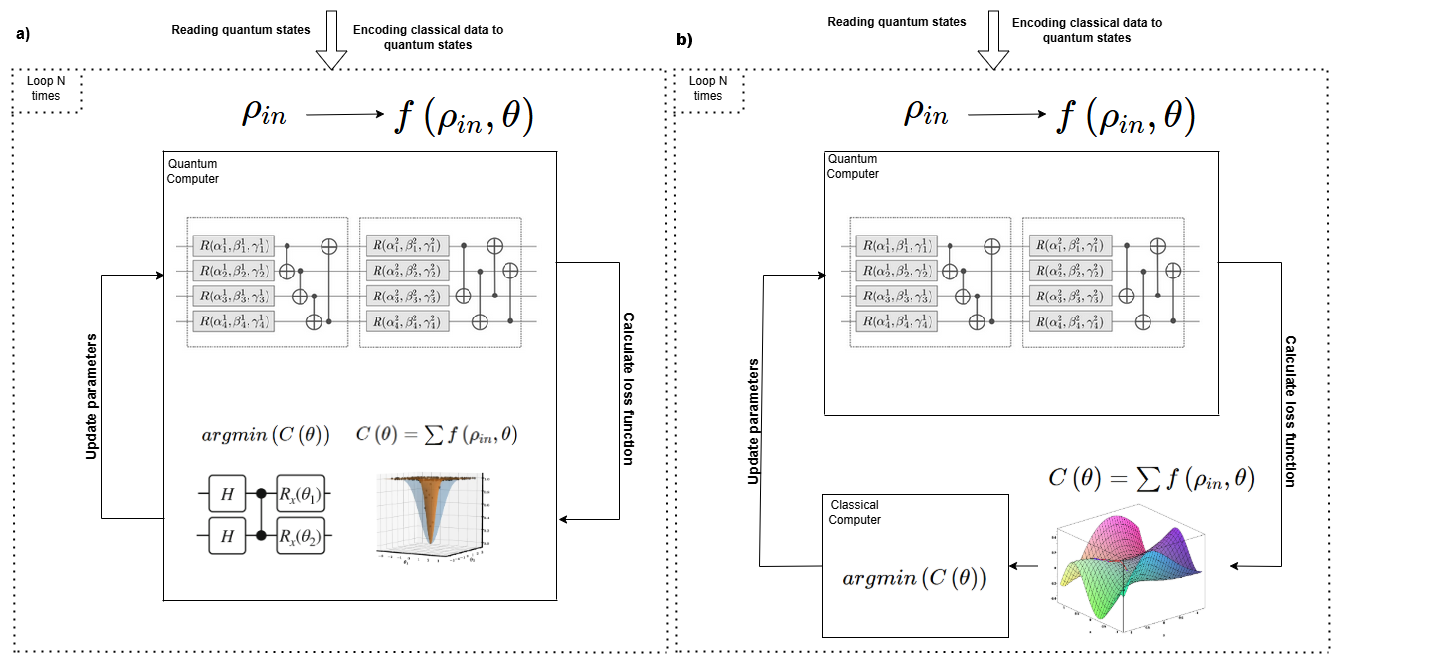}
{The two currently popular architectures of QML models are: a) A fully fledged QML architecture with computation and weight updates performed by quantum circuits. b) A hybrid classical QML architecture with computation optimization and weight updates performed by classical algorithms.\label{QDL}}
\subsubsection{QUANTUM GATES}
Like classical gates, quantum gates act on qubits and alter their states. Quantum gates can be single-qubit gates, such as X, Y, Z, S, T, R, Hadamard, etc., or two-qubit gates, such as CNOT, CX, CZ, SWAP, etc., and even three-qubit gates, such as CCX (Toffoli), CCZ, CSWAP (Fredkin), etc. Figure \ref{Had} illustrates the process of acting on the input state $| 0 \rangle$ after it is passed through the Hadamard gate.
\subsubsection{RANDOM QUANTUM CIRCUITS}
The combination of qubits with transformations through quantum gates creates quantum circuits. These quantum circuits are the main components in quantum algorithms such as Shor’s \cite{b37}, Grover’s \cite{b38}, and others, which have demonstrated superiority over classical algorithms. A RQC is a quantum circuit generated by randomly selecting gates according to a defined rule, aiming to create quantum transformations (unitaries) that are approximately Haar-random distributed in the Hilbert space \cite{b39}. RQC has demonstrated quantum supremacy in a recent Google study by generating a pseudo-random problem that is difficult to simulate on classical computers \cite{b32}. Figure \ref{RQC} illustrates an example of an RQC in which different lines refer to a different quantum gate applied at each transformation step in the circuit.  The results can be summarized in the formula:
\begin{equation}\dfrac{ \prod _{x\in S}\left| \langle x\right| \psi \rangle | ^{2}}{ \prod _{x\in S_{pcl}}\left| \langle x\right| \psi \rangle | ^{2}}\sim e^{m}\geq 1\label{eq3}\end{equation}
where $S$ is the quantum sample; $x$ is the measurement outcomes $\{0,1\}^n$ of the output state $| \psi \rangle$ of the RQC circuit; $\prod _{x\in S}\left| \langle x\right| \psi \rangle | ^{2}$ is the output sampling probability distribution of the RQC, and $\prod _{x\in S_{pcl}}\left| \langle x\right| \psi \rangle | ^{2}$ is the probability distribution a classical computer simulates. The formula indicates that a bit-string sampled from the output of an RQC $U$ with polynomial-size $n$ is essentially $e$ times more likely than a bit-string sampled from a classical algorithm.

\subsubsection{QUANTUM MACHINE LEARNING MODELS}
Quantum machine learning models, such as quantum classifier (QC) \cite{b7}, quantum convolutional neural networks (QCNN) \cite{b23}, quantum recurrent neural networks (QRNN) \cite{b47}, quantum vision transformers (QViT) \cite{b48}, and so forth, have demonstrated advantages in learning capabilities with only a limited amount of data compared to classical models \cite{b40}. However, in terms of resilience against adversarial attacks, they have not yet shown superiority over their classical counterparts \cite{b5}. Currently, quantum models are typically constructed using one of two approaches: fully fledged quantum or hybrid classical-quantum, where the main difference lies in whether the model’s parameters are optimized using quantum circuits or classical optimization algorithms, such as gradient descent or gradient-free methods.  

To process classical data $\rho_{in}$, specifically the image pixels used to train the model, an embedding process must be carried out. During this process, the numerical image pixels are encoded into qubits using one of the standard embedding methods, such as computational basis, amplitude, or angle rotation. The encoded quantum state can be represented as a function f with the classical data as its input parameter. Fig. \ref{QDL} illustrates the angle rotation embedding method, where the output function $f(\rho_{in}, \theta)$ takes the classical input parameter and the rotation angle $\theta$. The quantum state is then transformed through single-qubit and two-qubit gates, which creates entanglement among the qubits. This transformation process occurs within the Hilbert space, a massive space with dimensions of $2^n$, where $n$ is the number of qubits. The extent to which the quantum state is transformed depends on the depth of the quantum circuit, also known as the number of quantum layers. The output of the quantum circuit is collapsed into a classical bitstring based on measurements performed on the circuit. Because the quantum state in quantum mechanics is inherently a probability distribution, performing multiple measurements allows for an accurate diagnosis of the final quantum state. The quantum circuit's output, $\hat{y}$, is then fed into an optimizer to calculate the loss function $C(\theta)$ against the desired output $y$. This optimization can be done using a classical or quantum optimizer, depending on whether the model is a fully fledged quantum model or a hybrid classical-quantum model, as shown in Fig. \ref{QDL}. The following equation can express the entire learning process of the quantum circuit:
\begin{equation}
\begin{split}
\overline{u}\left( x,\theta ,\Xi \right) &= C\left( x,\theta ,\Xi \right) \\
&= \sum ^{n}_{i=0} \left\langle 0 \left| U_{emb}\left( x,\Xi \right)^{\dagger} O U_{\text{var}}\left( \theta \right) U_{emb}\left( x,\Xi \right) \right| 0 \right\rangle
\end{split}
\label{eq_QDL}
\end{equation}
where $C\left( x,\theta ,\Xi \right)$ is the desired output value of the quantum circuit, $n$ is the number of qubits, $U_{emb}\left( x,\Xi \right)$ represents the operator for encoding classical data into the quantum circuit, and $U_{\text{var}}\left( \theta \right)$ is the unitary operator that transforms the quantum state within the circuit to modify the $\theta$ parameter values.

\section{METHODOLOGY}
\Figure[t!](topskip=0pt, botskip=0pt, midskip=0pt)[width=2.0\columnwidth]{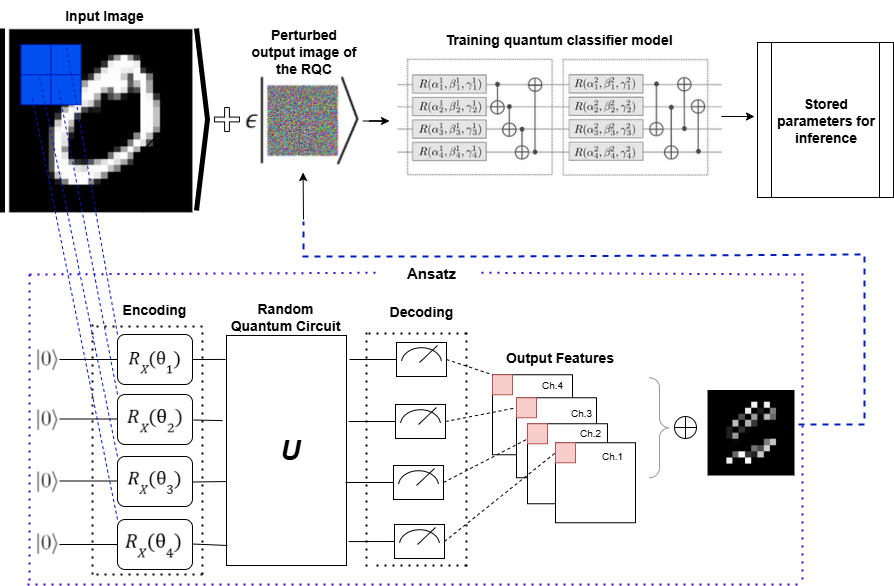}
{Overview of the proposed mechanism using a RQC to train QML models  against adversarial attacks. The magnified dashed rectangle illustrates the method of using a RQC to transform training images for the model. After transformation, the RQC with four qubits generates four channels, similar to the perturbed noise added in adversarial examples, then the data is used to train the  model.\label{ProposedMethod}}
Based on the superior properties of quantum computing, this research proposes a method utilizing RQCs to enhance the robustness of QML models against adversarial attacks. We leverage the transformation capability of RQCs in Hilbert space, which has been proven to approximate the Haar distribution, as a data augmentation method for the model. The proposed solution is presented in detail in the following sections.

\subsection{PROPOSED METHOD}
In this research, we exploit quantum patches to improve the robustness of QML models. Specifically, we transform the input image pixels using the RQC technique, where image patches are sequentially transformed, similar to a classical convolution filter. The RQC performs random transformations in the Hilbert space and generates layers of pseudo-noise, mimicking how adversarial attack methods affect legitimate images. It is expected that training QML with these pseudo-noise images will improve their robustness. To create a layer resistant to adversarial attacks, RQCs are used to transform input images into channels. Specifically, as illustrated in Figure \ref{ProposedMethod}, we generate four channels. The working mechanism of the RQC is analogous to a convolutional layer in a classical CNN and can be referred to as a quanvolutional layer \cite{b22}. The input data is encoded from pixels into the rotation angles of qubits. Currently, there are several encoding methods, among which the three most common are encoding into the computational basis, amplitude, or angle rotation. In this research, we chose angle rotation encoding because of its suitability for small local spatial data, such as the filter windows of the convolutional method. It doesn't require many qubits to encode an input image patch; in our experiment, we only needed four qubits. Furthermore, this method helps reduce computation time for embedding, resulting in a decrease in the total quantum processing time for the entire image. This decrease is particularly crucial in the NISQ era when applying quantum algorithms, especially since we are still often forced to use classical computers to simulate quantum computers.. After being encoded into four qubits, the data is transformed through a RQC unitary operator $U$. The circuit process, performed in the Hilbert space, generates new states for the qubits and follow the Haar probability distribution. 

After completing the transformation in the RQC, the state of the qubits is decoded through the measurement of the qubits. The measurement process collapses the qubit states into classical bits 0 and 1. Thus, the algorithm's time complexity depends only on the encoding and decoding processes. If compared to the convolution algorithm on a classical computer, its time complexity would be $O(n^2)$. After being transformed through the RQC quantum circuit, the data is combined with the original data to form a data layer used to counter adversarial attacks.

\subsection{METRICS}

\subsubsection{Adversarial Accuracy \& Attack Success Rate}
Adversarial accuracy (AA) is a metric used to evaluate the robustness of machine learning models—intense learning models—against adversarial attacks. While traditional accuracy is measured on clean, unaltered data, adversarial accuracy is assessed on data intentionally perturbed with slight, crafted noise designed to deceive the model. Suppose you have an adversarial dataset  $\left\{ x_{i}^{'},y_{i}\right\} _{i=1}^{N}$, where $x_{i}^{'}$ represents the input samples perturbed by adversarial noise, and $y_{i}$ denotes the corresponding accurate labels. The formula for calculating adversarial accuracy is given as follows:
\begin{equation}AA =\dfrac{1}{N}\sum ^{N}_{i=1}1[ f\left( x_{i}^{'}\right) = y_{i}] \label{eq4}\end{equation}
where $f(x_{i}^{'})$ denotes the model's prediction on the adversarial example  $x_{i}^{'}$.
Unlike adversarial accuracy, which assesses a model’s resilience, the Attack Success Rate (ASR) quantifies how successful an adversarial attack is against a deep learning model. Specifically, ASR represents the percentage of adversarial examples the model incorrectly classifies relative to their true labels. The ASR can be computed using the following formula:
\begin{equation}ASR =\dfrac{1}{N}\sum ^{N}_{i=1}1[ f\left( x_{i}^{'}\right) = y_{i} \wedge f(x_{i}^{'}) \neq y_{i}] \label{eq5}\end{equation}

\subsubsection{Lipschitz Bound}

\Figure[t!](topskip=0pt, botskip=0pt, midskip=0pt)[width=0.9\columnwidth]{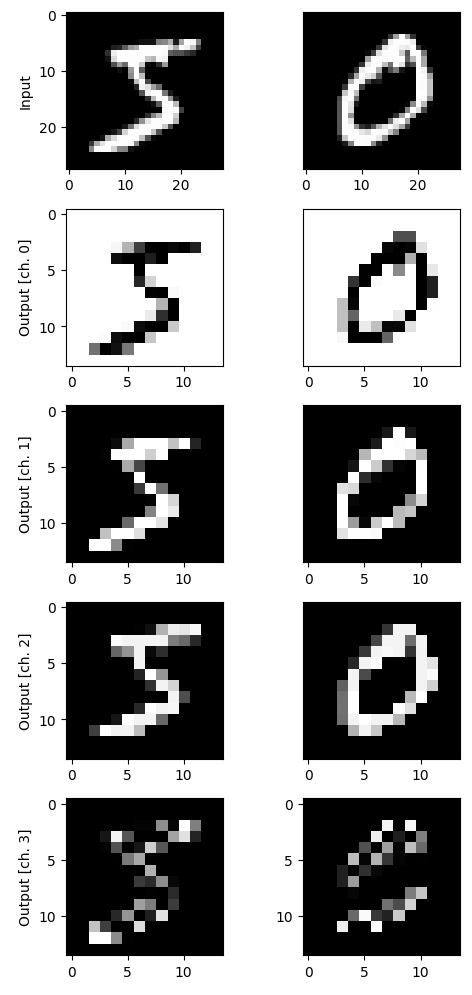}
{The output image, processed after the RQC transformation, consists of the original image and four additional channels.
\label{Quanvo}}

One of the strategies used to evaluate the robustness of a deep learning model—both classical and quantum—is to measure the model’s output response to varying inputs through the Lipschitz bound. Assuming the model is treated as a function $f$, the Lipschitz bound of $f$ is defined as the smallest constant $L$ that satisfies the following inequality:
\begin{equation}|| f\left( x\right) -f\left( y\right) \left\| \leq L\right\| x-y\|\label{eq6}\end{equation}
For a QML model, the Lipschitz bound \cite{b49} can be computed using the following formula:
\begin{equation}L_{\theta }=2\left\| M\right\| \sum ^{N}_{j=1}\left\| \omega_{j}\right\| \left\| H_{j}\right\|\label{eq7}\end{equation}
where and $\omega_{j}$ is fixed and $\theta$ is a trainable parameter.

\subsubsection{Average Fidelity}
Average fidelity is a metric used to quantify the difference between two quantum states \cite{b50}. Eq. \ref{eq8} presents the computation of this metric, where $\varepsilon$ denotes a trace-preserving quantum operation, U is an unitary gate, and $| \psi \rangle$ represents a d-dimensional quantum state. The average is taken over all possible quantum states, randomly and uniformly sampled according to the Haar distribution, normalized so $\int d\psi = 1$.
\begin{equation} \overline{F}\left( \varepsilon,U \right) \equiv \int d\psi \langle \psi \left| U^{\dagger} \varepsilon \left( \psi \right) U \right| \psi \rangle \label{eq8}\end{equation}

For this study, average fidelity is used to measure the quantum state before and after applying adversarial examples and processing through QML models. This measurement aims to evaluate the model’s ability to resist changes in the quantum state after quantum-enhanced defense techniques have been applied. In this research, we employ a method for calculating the average fidelity between two quantum states: one from the clean image and the other from the quantum state of the adversarial image \cite{b5}. This value is calculated immediately after the embedding of the qubit into the quantum circuit. The fidelity value $\overline{F}= \left| \langle \psi ^{adv}\right| \psi ^{leg}\rangle | ^{2}$ is calculated between $| \psi ^{leg}\rangle$ and $| \psi ^{adv}\rangle$, representing the quantum states after encoding the clean legitimate and adversarial examples.

\section{NUMERICAL SIMULATIONS}

\Figure[t!](topskip=0pt, botskip=0pt, midskip=0pt)[width=1.8\columnwidth]{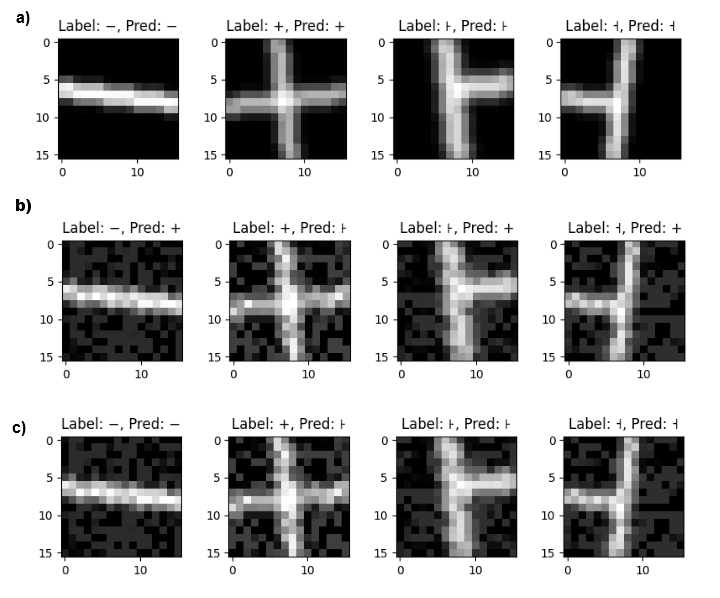}
{Experimental results on the Plus-Minus dataset: a) Model prediction results with clean data. b) Model prediction results without RQC transformation using perturbed data at $\epsilon =0.1$. c) Model prediction results after using RQC-enhanced solution with perturbed data at $\epsilon =0.1$.
\label{Plus-Minus}}

This section reports the experimental results of the proposed research solution's simulated attack and defense environment. The adversarial-resistant data layer is generated based on a hybrid classical-quantum CNN model \cite{b22}. In contrast, the models used for testing are a hybrid classical quantum model \cite{b7}. The datasets used in the experiments are Plus-Minus \cite{b6}, Modified National Institute of Standards and Technology (MNIST) \cite{b45}, Canadian Institute for Advanced Research (CIFAR-10) \cite{b46}, and CINIC-10 \cite{b17}. The programming language used for the experiments is Python, along with the Python open-source QML framework Pennylane \cite{b42}. We also utilized acceleration libraries, such as JAX \cite{b43} and Optax \cite{b44}, to improve the model's training speed. The QML models used in these experiments were trained on the REPACSS (Remotely-managed Power-Aware Computing Systems and Services) high-performance computing (HPC) infrastructure at Texas Tech University. We utilized compute nodes equipped with four NVIDIA H100 NVL GPUs, each providing 94 GB of HBM per GPU. The training time for each model was approximately 48 hours.

\subsection{SIMULATION SETUP}
The experiments compared the robustness of the non-RQC model against the RQC-applied model wherein the RQC-applied model is constructed similarly to the architecture proposed by Henderson \textit{et al.} \cite{b22}. These experiments are conducted using the following datasets: Plus-Minus, MNIST, CIFAR-10, and CINIC-10. Plus-Minus is a binary dataset consisting of 1,000 images, each sized 16×1 pixel with a 4-class classification output. Since this dataset is readily integrated into the Pennylane library, we chose to use it for initial method testing. MNIST is a dataset comprising 70,000 28×28 grayscale images of the digits 0-9, divided into 60,000 training images and 10,000 test images. CIFAR-10 consists of 60,000 32×32 color images across 10 classes (airplane, automobile, bird, cat, deer, dog, frog, horse, ship, truck), with 50,000 for training and 10,000 for testing. Finally, the CINIC-10 dataset is an extended version of CIFAR-10, featuring 60,000 32×32 RGB images, along with 210,000 images of the same size from the ImageNet dataset, divided into three equal subsets (each containing 90,000 images) for training, validation, and testing. Specifically, we used the MNIST dataset with classes 0 and 1 to compare the robustness against adversarial perturbations between our RQC-applied quantum classifier and an ordinary quantum classifier. Furthermore, we conducted experiments comparing the post-attack robustness of our quantum classifier before and after applying RQC using the CIFAR-10 dataset with two classes: Cat and Dog. This same process is then applied to the CINIC-10 dataset.

\begin{table*}[htbp]
\centering
\caption{The experimental results compare the effectiveness of the RQC-enhanced method on several standard datasets, including MNIST, CIFAR-10, and CINIC-10, using a quantum classifier model.}
\setlength{\tabcolsep}{3pt}
\begin{tabular}{|p{180pt}|p{40pt}|p{40pt}|p{70pt}|p{80pt}|}
\hline
\textbf{Dataset} & \textbf{AA (\%)} & \textbf{ASR (\%)} & \textbf{AA with RQC (\%)}  & \textbf{ASR with RQC (\%)} \\
\hline
 MNIST Clean Images  & 100 &    & 88 (-12)  &  \\
 \hline
 MNIST Adversarial Examples ($\epsilon$ = 0.1) &  &  22  &  & 25.38 (+3.58) \\
 \hline
 CIFAR-10 Clean Images & 48 &    & 45 (-3)  &  \\
 \hline
 CIFAR-10 Adversarial Examples ($\epsilon$ = 0.1) &  &  89.8  &   & 68.45 (\textbf{-21.35)} \\
 \hline
 CINIC-10 Clean Images & 57 &    & 47 (-10)  &  \\
 \hline
 CINIC-10 Adversarial Examples ($\epsilon$ = 0.1) &  &  94.23  &   & 78.68 (\textbf{-15.55)} \\
 \hline
\end{tabular}
\label{tab:ExpRes}
\end{table*}

To generate noisy image data that closely resembles the data produced by adversarial attack methods, we utilize the 'quanvolution' approach, where the RQC is the core of the algorithm \cite{b22}. The quantum circuit transforms the input image into n channels (in this experiment, n = 4), similar to how filters in a CNN create output channels. These channels contain noise analogous to the noise caused by adversarial attacks, due to the inherent effects of the quantum circuit. Fig. \ref{Quanvo} displays the output channels after the RQC circuit transformation for the input image. We consider this transformation to be a form of pseudo-noise that closely resembles the perturbed noise generated by adversarial attacks. For the QML model used to test the effectiveness of the method, we selected a circuit-centric quantum classifier \cite{b7}. The mechanism of the data-reuploading scheme encodes m × m input pixels into the latent space of the quantum classifier. The architecture of the circuit-centric quantum classifier, with its strongly entangling layers, allows this type of circuit to reach the 'wide corners of the Hilbert space'. Within Pennylane, we utilize the StronglyEntanglingLayers template with the output being PauliZ gates acting on four qubits. The classical cross-entropy loss function is employed to optimize the circuit parameters during the model training process.

\subsection{RESULTS and ANALYSIS}
In this section, we present the experimental results of our method on standard image datasets used in machine learning. Fig. \ref{Plus-Minus} illustrates how our method improves the robustness of QML models against adversarial attacks. We used a noise level of $\epsilon =0.1$  and experimented on the Plus-Minus dataset. The result shown in Fig. \ref{Plus-Minus} a) is the model's prediction when tested with clean images, where the model showed perfect accuracy, correctly predicting all four input samples. Fig. \ref{Plus-Minus} b) shows the model's prediction results when tested with adversarially perturbed images, where the model incorrectly predicted all four samples. Finally, Fig. \ref{Plus-Minus} c) displays the results when the model is retrained with the RQC transformation applied to the input images; in this case, the model correctly predicted three out of the four adversarial examples.

The experimental results, presented in Table \ref{tab:ExpRes}, demonstrate that for low-feature datasets, such as MNIST, the model's accuracy is relatively high, reaching 100\%. However, when applying RQC to the data and training the model, it did not make the model more robust, as the ASR actually increased from 22\% to 25.38\%. However, when the method is tested on high-feature datasets, such as CIFAR-10 and CINIC-10, the effectiveness of the solution is quite evident. With the CIFAR-10 dataset, the ASR decreased from 89.8\% to 68.45\%, while on the CINIC-10 dataset, it dropped from 94.23\% to 78.68\%. Thus, we can tentatively conclude that the noise generated by RQC is effective on high-feature datasets. This phenomenon can be explained by the fact that as the number of features increases, the noise generated by RQC is aggregated across more features, thereby affecting the model's generalization capability in object detection.

\subsection{ABLATION STUDY}

\Figure[t!](topskip=0pt, botskip=0pt, midskip=0pt)[width=0.99\columnwidth]{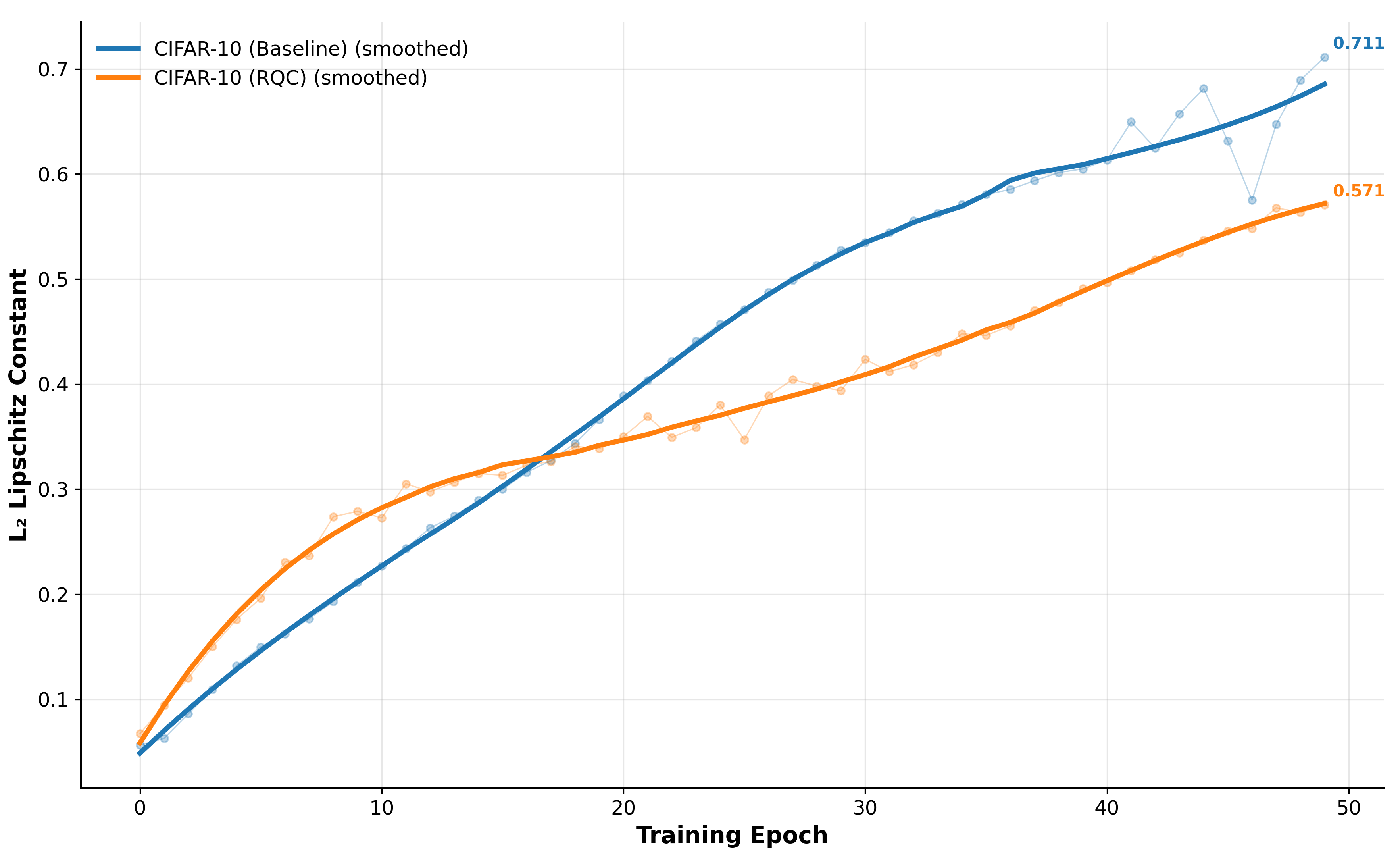}
{The change in the Lipschitz constant during the training process of the Baseline and RQC-enhanced models on the CIFAR-10 dataset. The orange line represents the results after applying the proposed RQC enhancement solution, which shows a lower Lipschitz value compared to the Baseline value. This lower value means that the output prediction value of the quantum classifier using RQC is not amplified when the input changes from a clean image to an adversarial image. This lower value implies that the RQC-enhanced model is more robust to noise. 
\label{fig:Lipschitz}}

To examine the actual influence of our solution's RQC on the quality of QML models, we performed an ablation study. In this experiment, we chose the Baseline as a quantum classifier retrained with perturbed images included in the training dataset, which we compared against our RQC-enhanced solution. This baseline method is one of the common approaches used to improve robustness against adversarial attacks.
We used two primary metrics for comparison: a classical metric, the Lipschitz bound, and a quantum metric, average fidelity. The experimental results from Fig. \ref{fig:Lipschitz} show that for the Lipschitz bound metric, the Lipschitz constant of the RQC-enhanced method initially increased faster in the early epochs. However, from epoch 16 onward, the increase slowed down, and its value became lower than that of the Baseline. At epoch 50, the Lipschitz value for RQC-enhanced was only 0.571, which is significantly lower than the 0.711 value obtained by the Baseline. This result demonstrates that the model shows better robustness to noise when trained with the RQC transformation.

\Figure[t!](topskip=0pt, botskip=0pt, midskip=0pt)[width=0.99\columnwidth]{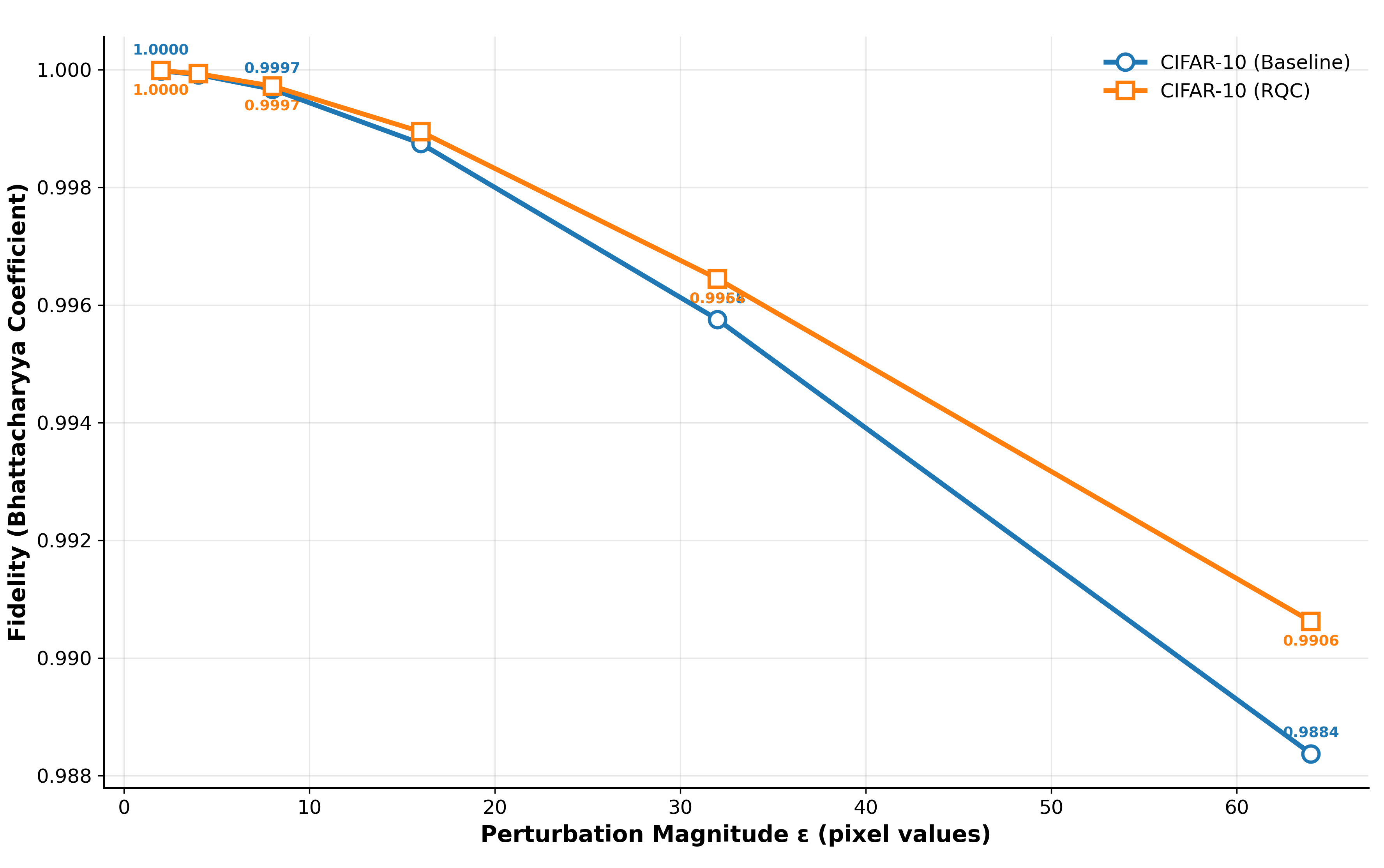}
{Comparison of the average fidelity of the output quantum states for the Baseline and RQC-enhanced models at the first epoch, with the epsilon perturbation magnitude gradually increasing on the CIFAR-10 dataset. The orange line represents the results after applying the RQC enhancement solution, which shows a higher average fidelity ($F$) compared to the Baseline as the noise intensity increases. A value of $F$ closer to 1 indicates that the quantum state output of the quantum circuit is less changed when the input quantum state is modified from a clean image to an adversarial image. This higher value implies that the RQC-enhanced model is more robust to noise. 
\label{fig:Fidelity}}

Regarding the output quantum state, as presented in the theoretical section on the average fidelity metric, this metric provides information about the transformation of the output quantum state between clean input image data and data with adversarial noise. If the output value is closer to 1, it indicates that the output quantum state is more stable with respect to the input state suggesting that the quantum model is less susceptible to perturbed noise. The experimental results from Fig. \ref{fig:Fidelity} show that at epoch 1, as the noise level gradually increases, the RQC-enhanced solution helps the quantum model maintain an average fidelity value that decreases more slowly and stays closer to 1 compared to the Baseline. This result confirms that the quantum model's robustness against adversarial noise is improved.

\section{CONCLUSIONS}

Enhancing the robustness of machine learning models against adversarial attacks is a critical research area, given the increasing prevalence of deep learning models in human life. In this study, we propose a method to improve the robustness of QML models against adversarial attacks by leveraging the transformation of quantum states within the Hilbert space effected by RQC. Experimental results demonstrate that the proposed method yields promising outcomes, reducing the adversarial attack success rate (ASR) for the quantum classifier model by 21.35\% on the CIFAR-10 dataset and 15.55\% on the CINIC-10 dataset.

However, the proposed method still has limitations, as it did not improve robustness against adversarial attacks for low-feature datasets, such as MNIST. Furthermore, the pseudo-noise generated by RQC also caused a decrease in the model's accuracy compared to when the RQC-enhanced method was not applied. This study uses the quantum classifier model to test the effectiveness of the proposed solution. To reinforce the already proven hypothesis that adversarial examples can transfer from a classical model to a quantum model \cite{b5}, as well as from one quantum model to another, we also experimented with a QCNN model. The experimental results also showed that the model's robustness improved after applying the solution; however, the improvement was not as pronounced as with the quantum classifier.

In the future, we will focus on exploring the combination of RQC-enhanced noise with the noise from decoherence within quantum circuits. We will investigate the influence of these two types of noise on the robustness of quantum models. The combination of these two noise sources may yield promising results for noise-based model improvement methods. Furthermore, studying the nature of the noise induced in machine learning models will help inform the appropriate selection of noise types for this method.

\section*{Appendix}
All source code used in the experimental section of the paper can be viewed at \href{https://github.com/billytran2404/QDAA-2025}{QDAA-2025}.

\bibliographystyle{ieeetr}
\bibliography{QDAA}

\phantomsection
\begin{IEEEbiography}[{\includegraphics[width=1in,height=1.25in,clip,keepaspectratio]{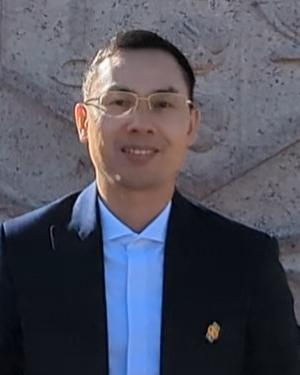}}]{BAN Q. TRAN} (Student Member IEEE) received the B.S degree in Electronics and Communication from Hanoi University of Science and Technology, Vietnam, in 2004 and a dual master’s degree in Computer Science from the University of Science and Technology of Hanoi, Vietnam, and La Rochelle Université, France, in 2021. He is currently pursuing a Ph.D. in Computer Science at Texas Tech University, Texas, USA. From 2017 to 2023, he was a senior lecturer and researcher in Computing Fundamentals at FPT University, Vietnam. His research focuses on new technologies such as Artificial Intelligence, Financial Technology, and Quantum Computing.
\end{IEEEbiography}

\phantomsection
\begin{IEEEbiography}[{\includegraphics[width=1in,height=1.25in,clip,keepaspectratio]{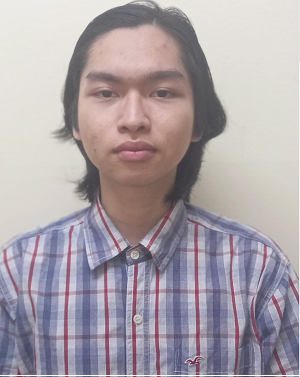}}]{CHUONG K. LUONG}  is currently pursuing a Bachelor's degree in Computer Science at FPT University, Hanoi, Vietnam. His academic interests include machine learning, 
computer vision, and software development.
\end{IEEEbiography}

\phantomsection
\begin{IEEEbiography}[{\includegraphics[width=1in,height=1.25in,clip,keepaspectratio]{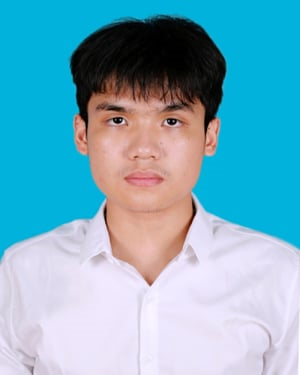}}]{VIET Q. NGUYEN}  is currently an undergraduate student pursuing a Bachelor's degree in Artificial Intelligence at FPT University, Hanoi, Vietnam. He is also a member of the SAP Innovation Lab, where he focuses on creating innovative AI solutions. His research interests include quantum computing, deep learning, and the application of AI to real-world problems.
\end{IEEEbiography}

\phantomsection
\begin{IEEEbiography}[{\includegraphics[width=1in,height=1.25in,clip,keepaspectratio]{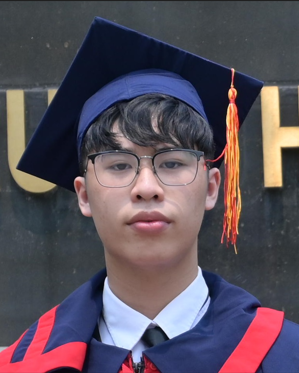}}]{DUONG M. CHU} born in 2004 in Hanoi, Vietnam, is studying to earn a Bachelor's in Artificial Intelligence at FPT University. He is also an active member of the SAP Innovation Lab at FPT University, where he engages in research and development initiatives. The research areas he is pursuing include Artificial Intelligence and Quantum Computing.
\end{IEEEbiography}

\phantomsection
\begin{IEEEbiography}[{\includegraphics[width=1in,height=1.25in,clip,keepaspectratio]{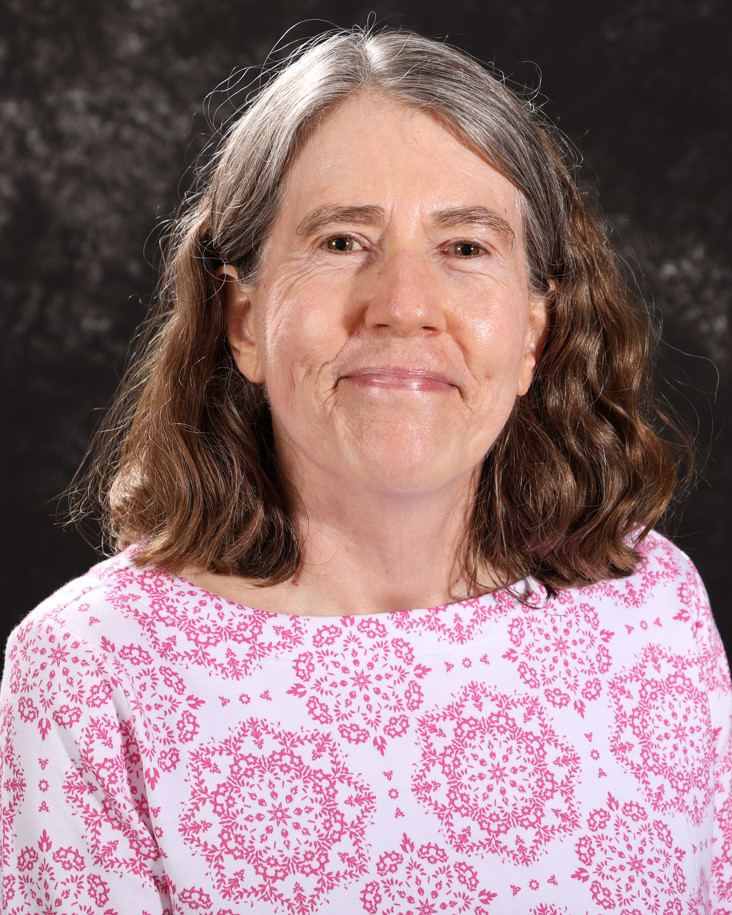}}]{Dr. SUSAN MENGEL} (Senior Member IEEE) received her Ph.D. from Texas A$\&$M University in 1990. She is an Associate professor at Texas Tech University where she is involved in NSF research projects for establishing a green computing cluster, computing cluster security, and in academic/industry collaborations. She has played strategic leadership roles in numerous transdisciplinary projects involving the delivery of innovative software and data models in sleep management, student retention and advising, computer education, nutrition, speech therapy, cardiovascular disease, and cybersecurity. She helped to establish the Master's in Software Engineering degree program at Texas Tech University, served as Associate Editor for Computing for the IEEE Transactions on Education, served on the Steering Committee of the ACM/IEEE Computing Curriculum, and served as the Outreach Chair FY19 of the SWE Outreach Committee. She currently serves as Associate Chair on the Texas Tech Institutional Review Board for the Protection of Human Subjects, is the Undergraduate Program Coordinator for the Department of Computer Science at Texas Tech, and is a faculty advisor for the TTU Collegiate Chapters of SWE and Women in High Performance Computing.
\end{IEEEbiography}

\EOD

\end{document}